# A Bayesian Framework for the Estimation of the Single Crystal Elastic Parameters from Spherical Indentation Stress-Strain Measurements

Andrew Castillo[1] and Surya R. Kalidindi[1*]

[1]George W. Woodruff School of Mechanical Engineering, Georgia Institute of Technology, Atlanta, GA 30332, USA

**Abstract**

This paper presents a two-step Bayesian framework for the estimation of the intrinsic single crystal elastic stiffness parameters from the measurements of spherical indentation stress-strain responses in multiple individual grains of a polycrystalline sample, whose crystal lattice orientations have been measured using electron back-scattered diffraction technique. The first step requires the establishment of the functional dependence of the indentation elastic modulus given the lattice orientation and the intrinsic single crystal elastic stiffness parameters. Previous efforts for this step required a large database of computationally expensive finite element (FE) simulations in order to establish this function with adequate accuracy. In this paper, it is shown that the introduction of a Bayesian framework can greatly reduce the number of simulations necessary to establish this function, while introducing practically useful measures of uncertainty which can guide the selection of specific additional simulations that are expected to best improve the predictive accuracy of the function. The second step involves a Markov Chain Monte-Carlo (MCMC) sampling of the distribution of possible values for the single crystal elastic stiffness parameters based on a given set of experimentally measured elastic indentation moduli in individual grains of different lattice orientations. This second step is accomplished by calibrating the available experimental data to the function established in the first step. This novel framework is presented and demonstrated in this paper for an as-cast cubic polycrystalline Fe-3% Si sample and a hexagonal polycrystalline commercially pure (CP-Ti) titanium sample.

## 1. Introduction

Continued development and application of physics-based multiscale materials models is largely hampered by the lack of protocols for reliably estimating the intrinsic material properties at the microscale (e.g., the grain-scale properties in modeling of polycrystalline materials). In recent years, instrumented indentation techniques have been demonstrated to be capable of providing consistent and reliable measurements at the lower length scales (up to submicron length scales) [1-5]. Although small-scale mechanical measurements are now quite reliable, it has not been a straightforward process to extract the intrinsic material properties from such measurements. As specific examples, one would hope to estimate the values of the single crystal elastic constants and the critical resolved shear strengths from the instrumented nanoindentation measurements. Reliable and robust protocols for addressing this gap are emergent [6-8].

Currently employed strategies for extracting intrinsic material properties from indentation tests have generally involved the calibration of physics-based finite element (FE) models of these tests to the corresponding set of experimental measurements [6, 9-11]. In this regard, it has been pointed out in recent work [7] that these protocols are much more robust when the calibration is attempted in the form of the normalized indentation stress-strain curves as opposed to directly matching the load-displacement curves. This is mainly because the initial elastic response and the elastic-plastic transition occur over a very short early portion of the load-displacement curve that is not easily identified and isolated, resulting in a very high sensitivity of the extracted values of the intrinsic material properties to small changes in the calibration procedures.

The calibration of the FE simulated indentation stress-strain curves to the experimentally measured indentation stress-strain curves for any selected material system essentially involves solving an inverse problem. In other words, the guessed values of the intrinsic material properties of interest become inputs to the FE simulations. Typically, one has to search over a large multidimensional space to find the best-fits between the FE predictions and the measurements. The main challenge comes from the high computational expense of FE simulations of the indentation experiments. It should be noted that establishing each data point on the FE predicted indentation stress-strain curve needs the simulation of a suitable unloading segment [7], and this drives up the cost of the simulation significantly. Given all of the complexity described, the only logical path forward is to establish a reduced-order model for the FE simulations of the indentation test, and to use the reduced-order model in solving the inverse problem described above. In recent work [9], we have formalized this approach as a two-step process: (1) establishing a reduced-order model calibrated to FE simulations of indentations that takes the relevant intrinsic material properties as inputs and predicts indentation properties (defined suitably on an indentation stress-strain curve), and (2) the extraction of the intrinsic material properties from the available measurements (typically performed on grains of different orientations in a polycrystalline sample) through calibration with the reduced-order model established in step (1). The second step described above typically involves the solution to an optimization problem (i.e., minimizing the difference between the measurements and the predictions from the reduced-order model). The viability of this two-step protocol for extracting the values of the single crystal elastic constants and the critical resolved shear strengths in Fe-3%-Si has been demonstrated in recent work [7, 9].

The main difficulty with the two-step protocol described above lies in building the reduced-order model (i.e., step (1)). Because of the need to cover a large space (for example for extracting single crystal elastic constants, the input space of interest is the product space spanning all combinations of the single crystal elastic constants, $C_{11}, C_{12}, C_{44}$, and all possible grain orientations), one needs to generate a large amount of the FE simulation data in order to establish a high-fidelity reduced-order model. The difficulty of this task is amplified significantly in dealing with hcp crystals, where the numbers of the intrinsic properties is significantly larger (for example, modeling the elastic deformation in hcp crystals requires specification of five independent single crystal elastic constants). In prior work [9], the reduced-order models were built using standard regression approaches. Although these regression approaches produced excellent results, they do not scale well to problems with larger numbers of the intrinsic properties (because of the need to generate a large amount of data spanning the entire input domain).

The primary goal of this paper is to demonstrate the utility of Bayesian strategies for (i) optimizing the reduced-order model building effort involved in step (1), and (ii) providing estimates of the desired intrinsic material parameters (single elastic constants specifically) with uncertainty measures from available experimental data (spherical indentation measurements). Towards these goals, we will develop and present a Bayesian inference framework for both steps of the two-step protocol described above. Bayesian inference has been instrumental in model-building tasks with limited amount of data [12-15]. The adoption of a Bayesian inference framework for the extraction of the intrinsic material properties from indentation measurements offers the following main advantages: (i) it is expected to dramatically reduce the number of FE simulations needed to produce the reduced-order model generated in step (1), and (ii) it provides a much more rigorous quantification of the uncertainty in the estimates of the intrinsic material properties obtained in step (2), while accounting for the uncertainty in the measurements as well as other sources. In this paper, we first develop the framework, and subsequently demonstrate its application to the extraction of single crystal elastic properties in selected cubic and hexagonal metals.

## 2. New Bayesian Inference Framework for the Estimation of Intrinsic Material Properties from Indentation Measurements

Let $\boldsymbol{c}$ denote the set of intrinsic material properties to be established. For cubic crystals, this represents the set of three elastic constants, i.e., $\boldsymbol{c} = \{C_{11}, C_{12}, C_{44}\}$. Let $\boldsymbol{P}$ denote an available set of observations of the indentation properties corresponding to the set of crystal orientations $\boldsymbol{G}$. This set of observations could come from either FE simulations or the physical experiments. We shall note the source of the data using subscripts *sim* and *exp* on these variables. Furthermore, in the notation employed in this paper, a set of values for a variable is denoted by an upper-case symbol, while an individual element of the set is denoted by its non-bold counterpart. As an example, a single value of the indentation property will be denoted by $P$. Furthermore, a collection of variables is also denoted by bold symbols. As an example, a single crystal orientation would be denoted by $\boldsymbol{g}$, as it denotes a set of three Bunge-Euler angles [16]. However, a collection of grain orientations would be represented by $\boldsymbol{G}$. Likewise, a single set of intrinsic material parameters is denoted by $\boldsymbol{c}$, whereas a collection of the sets of intrinsic material parameters is denoted $\boldsymbol{C}$. Employing this notation, the central tasks in the two-step protocol developed in this work are the following:

1. Establish a reduced-order model that takes given values of $\boldsymbol{c}$ and $\boldsymbol{g}$ and predicts the indentation property of interest, $P = \hat{P}(\boldsymbol{c}, \boldsymbol{g})$, while employing Bayesian inference in building the reduced-order model. In other words, given the previously aggregated set of simulation data $\{\boldsymbol{P}_{sim}, \boldsymbol{G}_{sim}\}$, determine the new inputs for the FE simulation that would yield the best improvements in the reliability of the reduced-order model being built.
2. Given the reduced-order model built in step (1) and a set of experimental observations $\{\boldsymbol{P}_{exp}, \boldsymbol{G}_{exp}\}$ from a given polycrystalline sample, establish the posterior distribution on $\boldsymbol{c}$ for the sample. It is noted that the indentation properties are measured by the spherical indentation protocols mentioned earlier, while the orientations are measured using electron back-scattered diffraction (EBSD) techniques [17].

Prior experimental work [37] in single-phase polycrystalline metals has focused on exploring the dependence of indentation modulus on the lattice orientation of the indented grains (i.e., individual crystals). These findings were verified by suitable FE simulations [9]. Recently, a reduced-order model which captures the dependence of indentation modulus on both orientation and an arbitrary set of intrinsic material parameters has been established from FE simulations. The mathematical form of the reduced-order model for the present application is adopted from this prior work [9] as

$$P = \hat{P}(\boldsymbol{c}, \boldsymbol{g}) \approx \sum_{l=0}^{L} \sum_{m=1}^{M(l)} \sum_{\boldsymbol{q}}^{\boldsymbol{Q}} A_l^{mq} \mathrm{K}_l^m(\boldsymbol{g}) \widetilde{\mathrm{P}}^q(\bar{\boldsymbol{c}}) \tag{1}$$

$$\bar{c}_j = \frac{2c_j - c_j^{min} - c_j^{max}}{c_j^{max} - c_j^{min}} \tag{2}$$

where $\mathrm{K}_l^m(\boldsymbol{g})$ denote the symmetrized Surface Spherical Harmonics basis over the relevant orientation space of interest, and $\widetilde{\mathrm{P}}^q(\ )$ denote a multivariate Legendre polynomial product basis. In other words, one can express $\widetilde{\mathrm{P}}^q(\bar{\boldsymbol{c}}) = \mathrm{P}^{q_1}(\bar{c}_1)\mathrm{P}^{q_2}(\bar{c}_2)\ldots\mathrm{P}^{q_R}(\bar{c}_R)$, where $\boldsymbol{q} = (q_1, q_2 \ldots q_R)$ forms a multi-index array, each element of which is a nonnegative integer allowed to vary from 0 to the selected maximum degree, $Q$, i.e., $q_j \in [0, Q]$. The use of Legendre polynomials provides an orthonormal basis over the range [-1,1], for which each of the elastic constants are rescaled in accordance to Eq. (2), where $c_j^{max}$ and $c_j^{min}$ are the maximum and minimum values of the $j$-th elastic constant under consideration. In Eq. (1), $m$ and $l$ index the surface spherical harmonic basis where $M(l)$ enumerates the spherical harmonics that implicitly reflect the crystal symmetries of interest [16, 18]. The integers $Q$ and $L$ denote the truncation levels adopted in the use of Eq. (1). It is emphasized here that the model form used in Eq. (1) denotes a Fourier representation using an orthonormal basis that has been previously shown to produce compact representations for mechanical responses of crystalline solids [7, 9, 19-22]. One of the central features of a Fourier representation is that the Fourier coefficients $A_l^{mq}$ are completely independent of each other. The goal of the reduced-order modeling task here is to estimate the values of $A_l^{mq}$, expressed in a vector notation as $\boldsymbol{A}$, from the sparse amount of available data, as it is being generated from the expensive FE simulations. Even more importantly, our goal is to drive the model building in an optimal way by identifying the specific set of inputs for the next FE simulation such that it maximizes the improvement to the reduced-order model being built.

## 2.1 Building the Reduced-Order Model

The reduced-order model (see Eq. (1)) needs to be built such that it makes good predictions for the indentation modulus over a large domain of input parameters $(\boldsymbol{c}, \boldsymbol{g})$. Given the large domain of the input parameters (e.g., covering the range of values for the three independent parameters defining cubic elasticity and the two independent parameters defining the indentation direction in the crystal reference frame) and the high cost of executing a FE simulation for generating each data point, it is highly desirable to explore Bayesian regression approaches for estimating the unknown Fourier coefficients in Eq. (1). Let the corresponding sets of intrinsic parameters, $\boldsymbol{c}$, used as inputs to simulations be denoted as $\boldsymbol{C}_{sim}$. The data generated from FE simulations will be denoted $\{\boldsymbol{P}_{sim}, \boldsymbol{C}_{sim}, \boldsymbol{G}_{sim}\}$ following the notation introduced earlier.

Bayesian approaches treat model parameters (e.g., Fourier coefficients in Eq. (1)) as stochastic variables exhibiting a distribution of values. Most importantly, Bayes' theorem allows one to update the distributions for the model parameters given new data (i.e., observations) and is commonly expressed as

$$P(A|D) = \frac{P(D|A)P(A)}{P(D)} \quad (3)$$

where $P(A)$ denotes the prior belief (expressed as a distribution) on the values of the unknown model parameters, $P(D|A)$ denotes the likelihood of sampling the observations $D$ for specified values of the model parameters, and $P(A|D)$ denotes the posterior (updated) belief on the values of the unknown model parameters given the observations $D$. The denominator $P(D)$ in Eq. (3) is generally referred as the probability of the evidence, and is often difficult to establish. However, it mainly serves as a normalization factor for the posterior distribution. Since the distributions are often defined with known normalization factors, it is often possible to skip the evaluation of $P(D)$ in practical implementations of the Bayes' rule described in Eq. (3) [23].

It is expedient to treat the distributions associated with all the stochastic variables in Eq. (3) as normal (i.e., Gaussian) distributions. As a specific example, the i[th] observed value of the indentation modulus is modeled as being generated from a deterministic model, with added stochastic noise, as

$$P_i = \hat{P}_i(\boldsymbol{A}, \boldsymbol{c}_i, \boldsymbol{g}_i) + \varepsilon_i, \quad \varepsilon_i \sim \mathcal{N}(0, \beta^{-1}) \tag{4}$$

where $\mathcal{N}(0, \beta^{-1})$ denotes a normal distribution with a zero mean and a variance of $\beta^{-1}$. Note that the stochastic noise is assumed to be independent of location in the parameter space, i.e., homoscedastic. The likelihood for a set of $N$ independently observed indentation moduli can be established using the product rule as

$$p(\boldsymbol{P}_{sim}|\boldsymbol{A}, \boldsymbol{C}_{sim}, \boldsymbol{G}_{sim}, \beta) = \prod_i^N p(P_i|\boldsymbol{A}, \boldsymbol{c}_i, \boldsymbol{g}_i, \beta) \tag{5}$$

As noted earlier, the model parameters $\boldsymbol{A}$ are also treated as stochastic variables. The prior belief on these variables is assumed to be specified by a normal distribution with a zero mean and a large variance of $\alpha^{-1}$ as

$$p(\boldsymbol{A}|\alpha) \sim \mathcal{N}(0, \alpha^{-1}\boldsymbol{I}) \tag{6}$$

The application of Bayes' rule (Eq. (3)) to the problem at hand results in

$$p(\boldsymbol{A}|\boldsymbol{P}_{sim}, \boldsymbol{C}_{sim}, \boldsymbol{G}_{sim}, \alpha, \beta) = \frac{p(\boldsymbol{P}_{sim}|\boldsymbol{A}, \boldsymbol{C}_{sim}, \boldsymbol{G}_{sim}, \beta)\, p(\boldsymbol{A}|\alpha)}{p(\boldsymbol{P}_{sim}|\boldsymbol{C}_{sim}, \boldsymbol{G}_{sim}, \alpha, \beta)} \tag{7}$$

where $p(\boldsymbol{A}|\boldsymbol{P}_{sim}, \boldsymbol{C}_{sim}, \boldsymbol{G}_{sim}, \alpha, \beta)$ denotes the posterior (updated) distribution on the model parameters. The denominator in Eq. (7) reflects the probability of the observed outcomes irrespective of the model parameters $\boldsymbol{A}$ chosen, and can be described by the marginalization of the likelihood with respect to the model parameters as

$$p(\boldsymbol{P}_{sim}|\boldsymbol{C}_{sim}, \boldsymbol{G}_{sim}, \alpha, \beta) = \int_{\boldsymbol{A}} p(\boldsymbol{P}_{sim}|\boldsymbol{A}, \boldsymbol{C}_{sim}, \boldsymbol{G}_{sim}, \beta)\, p(\boldsymbol{A}|\alpha) d\boldsymbol{A} \tag{8}$$

In a fully Bayesian approach, the precision parameters, $\alpha, \beta$, may also be treated as stochastic variables [24]. This allows for a separate application of Bayes' theorem expressed as

$$p(\alpha, \beta|\boldsymbol{P}_{sim}, \boldsymbol{C}_{sim}, \boldsymbol{G}_{sim}) \propto p(\boldsymbol{P}_{sim}|\boldsymbol{C}_{sim}, \boldsymbol{G}_{sim}, \alpha, \beta) p(\alpha, \beta) \tag{9}$$

Alternately, one can use point estimates from the maximization of the likelihood in Eq. (9), denoted as $\hat{\alpha}, \hat{\beta}$. This is equivalently interpreted as the maximization of the evidence of the observed data in Eq. (8) [25]. With this approach, the posterior distributions of model coefficients in Eq. (7) can be solved analytically (while assuming normal distributions for the various variables involved) [25-27]. The updated posterior distribution computed using the approach described above is generally expected to be sharper (i.e., lower variance) compared to the prior belief.

Obviously, the available observations may not produce a posterior distribution that is sharp enough (i.e., the uncertainty associated with the posterior is still too high for a given application). In such cases, one needs to examine carefully where one should produce additional data points (i.e., new observations) in order to maximize the sharpening of the posterior distributions. The general approach to solving this problem (i.e., identifying the new data points exhibiting the maximum potential for improving the model accuracy and reliability) involves making predictions for new inputs, and identifying the specific inputs that exhibited the highest variance (i.e., uncertainty) in their predictions as the locations where new observations should be generated [28, 29]. This kind of a rational approach for deciding where to generate new data points is critical for situations where data generation is expensive (as is the case with the FE simulations of the spherical indentation for the present case study). The predictions for new inputs are obtained by the marginalization over the posterior distribution of the model parameters as

$$p(P|\mathbf{c},\mathbf{g},\mathbf{P}_{sim},\mathbf{C}_{sim},\mathbf{G}_{sim},\hat{\alpha},\hat{\beta}) = \int_A p(P|\mathbf{A},\mathbf{c},\mathbf{g},\hat{\beta})\, p(\mathbf{A}|\mathbf{P}_{sim},\mathbf{C}_{sim},\mathbf{G}_{sim},\hat{\alpha},\hat{\beta})\, d\mathbf{A} \quad (10)$$

where $(\mathbf{c},\mathbf{g})$ denote the new inputs. Therefore, the specific set of inputs which exhibit the highest variance for the prediction can be readily identified. Once the set of inputs are identified, and corresponding FE simulation performed, the next step is updating the distribution of model coefficients with the newly acquired observation. The update step to the distribution of the model coefficients is natural using a Bayesian framework in the sense that any knowledge acquired previously can be incorporated through the prior.

$$\begin{aligned} p_{N+1}(\mathbf{A}|\mathbf{P}_{sim},\mathbf{C}_{sim},\mathbf{G}_{sim},\hat{\alpha},\hat{\beta}) \\ \propto p_{N+1}(\mathbf{P}_{sim}|\mathbf{A},\mathbf{C}_{sim},\mathbf{G}_{sim},\hat{\beta})\, p_N(\mathbf{A}|\mathbf{P}_{sim},\mathbf{C}_{sim},\mathbf{G}_{sim},\hat{\alpha},\hat{\beta}) \end{aligned} \quad (11)$$

The posterior distribution of the parameters can continually be updated as incoming data is sequentially added by setting the prior as the previously inferred posterior distribution of model coefficients as shown in Eq. (11). Updates to the posterior distribution of model coefficients are performed until sufficient model convergence and prediction performance is attained. Model convergence is determined through the change in values of the model coefficients and parameters as data is added. Model performance is evaluated through various error metrics such as the leave-one-out-cross-validation (LOOCV) error [25, 26]. Building the reduced-order model and critically evaluating its reliability and robustness completes the first step of the two-step protocol. It should be noted that this is intended to be performed only once for a given class of materials.

**2.2 Estimating Intrinsic Material Properties from Indentation Measurements**

For the second step of the protocol, our goal is to employ the reduced-order model built in the first step together with indentation measurements obtained from a given sample to estimate its intrinsic material properties. Let $\{P_{exp}, G_{exp}\}$ denote such experimental measurements. The posterior distribution for the intrinsic material properties can be sampled from yet another application of the Bayes' rule as

$$p(c|A, P_{exp}, G_{exp}, \sigma) \propto p(P_{exp}|A, c, G_{exp}, \sigma)p(c) \qquad (12)$$

where $A$ denotes the parameters in the reduced-order model built in the first step. Although point estimates can be obtained by maximizing the likelihood in Eq. (12), in the spirit of building a robust framework capable of accounting for various sources of uncertainty, we have decided to pursue the computation of the posterior distribution on the intrinsic material properties through sampling techniques. In order to sample from the posterior distribution defined in Eq. (12), we need to establish the likelihood of the set of experimental observations. A likelihood can be constructed by assuming that the experimental observations (i.e., data points) are independent and normally distributed, i.e., the experimental data points are observations drawn from normal distributions with means estimated by the reduced order model and variances, $\sigma$, estimated from the experimental data of the measured indentation property at M grain orientations [23, 30-32]. This likelihood is expressed as

$$p(P_{exp}|A, c, G_{exp}, \sigma) = \prod_{i}^{M} \mathcal{N}(P_{i_{exp}}|\hat{P}(A, c, g_{i_{exp}}), \sigma_i) \qquad (13)$$

The evaluation of the likelihood described in Eq. (13) is performed using the reduced-order model, $\hat{P}(A, c, g)$, built in the first step of the two-step protocol. In this work, the sampling from the posterior distribution of intrinsic material parameters (Eq. (12)) is accomplished using a Monte Carlo Markov Chain (MCMC). The goal of MCMC is to generate a Markov Chain which indirectly samples from the posterior distribution of interest as long as the number of samples drawn is very large. The Markov Chain is generated by the acceptance and rejection of a large number of transitions through the space of intrinsic material parameters based on an acceptance probability. In practice, a class of algorithms have been developed in order to define these transitions and are referred as Metropolis-Hastings algorithms [24]. In this work, Single Component Metropolis Hastings (SCMH) is applied, which considers component wise transitions [33]. In the algorithm below for a given step $t$, partial updates are performed for the sample $\boldsymbol{c}_t$ for each component $j$ until all components are updated.

The basic steps for the implementation of the SCMH algorithm are as follows:

1. Initialize a starting point, $\boldsymbol{c}_0$, using the best available information
2. Sample transition, $\boldsymbol{c}^*$, from a proposal distribution $q_j(*)$ for an update of component $j$. If $t$ is a new step, initialize $\boldsymbol{c}_t = \boldsymbol{c}_{t-1}$ where $\boldsymbol{c}_t$ will be subjected to partial updates (one component at a time). Mathematically, one can express this as

    $$\boldsymbol{c}^* \sim q_j(\boldsymbol{c}|\boldsymbol{c}_t)$$

    where $q_j(*)$ proposes $\boldsymbol{c}^*$ differing from $\boldsymbol{c}_t$ in component $j$, sampled from a normal distribution with mean $c_t^j$ and variance $v_j^2$

    $$c^{j^*} \sim \mathcal{N}(c^j | c_t^j, v_j^2)$$

3. Calculate the acceptance probability of transition, $\alpha(*)$

    $$\alpha(\boldsymbol{c}^*|\boldsymbol{c}_t) = \min(1, \frac{p(\boldsymbol{c}^*|\boldsymbol{A}, \boldsymbol{P}_{exp}, \boldsymbol{G}_{exp}, \boldsymbol{\sigma})q_j(\boldsymbol{c}_t|\boldsymbol{c}^*)}{p(\boldsymbol{c}_t|\boldsymbol{A}, \boldsymbol{P}_{exp}, \boldsymbol{G}_{exp}, \boldsymbol{\sigma})q_j(\boldsymbol{c}^*|\boldsymbol{c}_t)})$$

    $$= \min(1, \frac{p(\boldsymbol{P}_{exp}|\boldsymbol{A}, \boldsymbol{c}^*, \boldsymbol{G}_{exp}, \boldsymbol{\sigma})p(\boldsymbol{c}^*)}{p(\boldsymbol{P}_{exp}|\boldsymbol{A}, \boldsymbol{c}_t, \boldsymbol{G}_{exp}, \boldsymbol{\sigma})p(\boldsymbol{c}_t)})$$

4. Update Chain (accept/reject proposed transition)
    a. Draw a sample, $r$, from a standard uniform distribution
    b. If $\alpha > r$

    $$\boldsymbol{c}_t = \boldsymbol{c}^*$$

5. Repeat steps (2-4) until all components of $\boldsymbol{c}_t$ are updated, then proceed to a new step.

While the probability of a proposed transition is described by the proposal distribution $q_j(*)$, the probability of accepting the transition is given by $\alpha(*)$. By assuming a flat prior for $p(\boldsymbol{c})$, the acceptance probability of a proposed transition is completely specified by the posterior probability of the states evaluated within a normalizing constant using Eq. (13) [24, 34]. The variances of the proposal distributions $v_j^2$ are tuned during the "burn-in" period in order to meet an acceptance rate around ~0.23. Ensuring the acceptance rate lies around 0.23 has been shown to provide efficient convergence of the Markov chain for gaussian posteriors [35]. All of the computations described above were realized using functions readily available in MATLAB [36].

## 3. Case Study: Cubic Polycrystals

### 3.1 Problem Statement

For our first case study, we revisit the extraction of the single crystal elastic constants $\{C_{11}, C_{12}, C_{44}\}$ of the bcc metal Fe 3%-Si, which was previously attempted using standard regression techniques. In the previous study [13], a total of 2286 simulations were needed to establish a high-fidelity reduced-order model in the first step of the two-step protocol. The simulated database consisted of the indentation modulus corresponding to 300 distinct sets of cubic stiffness constants within the domain 50 GPa $\leq C_{11} \leq$ 250 GPa, 40 GPa $\leq C_{12} \leq$ 150 GPa and 15 GPa $\leq C_{44} \leq$ 120 GPa across 9 orientations selected within the fundamental zone of the relevant orientation space [9]. We note, these ranges encompass a vast number of cubic metals [40]. It is anticipated that the proposed Bayesian framework will need significantly less number of FE simulations to adequately capture the FE predicted indentation modulus within the same parameter space in a robust reduced-order model.

### 3.2 Model Building Process

The Bayesian model building process enables sequential design strategies through the identification of high value simulations which will best improve the predictive capability of the model. Since a database of simulations is already available, simulations are treated as "unseen" and are sampled based on the determined utility of performing the simulation. Before beginning the sequential design process, an initial set of simulations must be performed to establish an initial model.

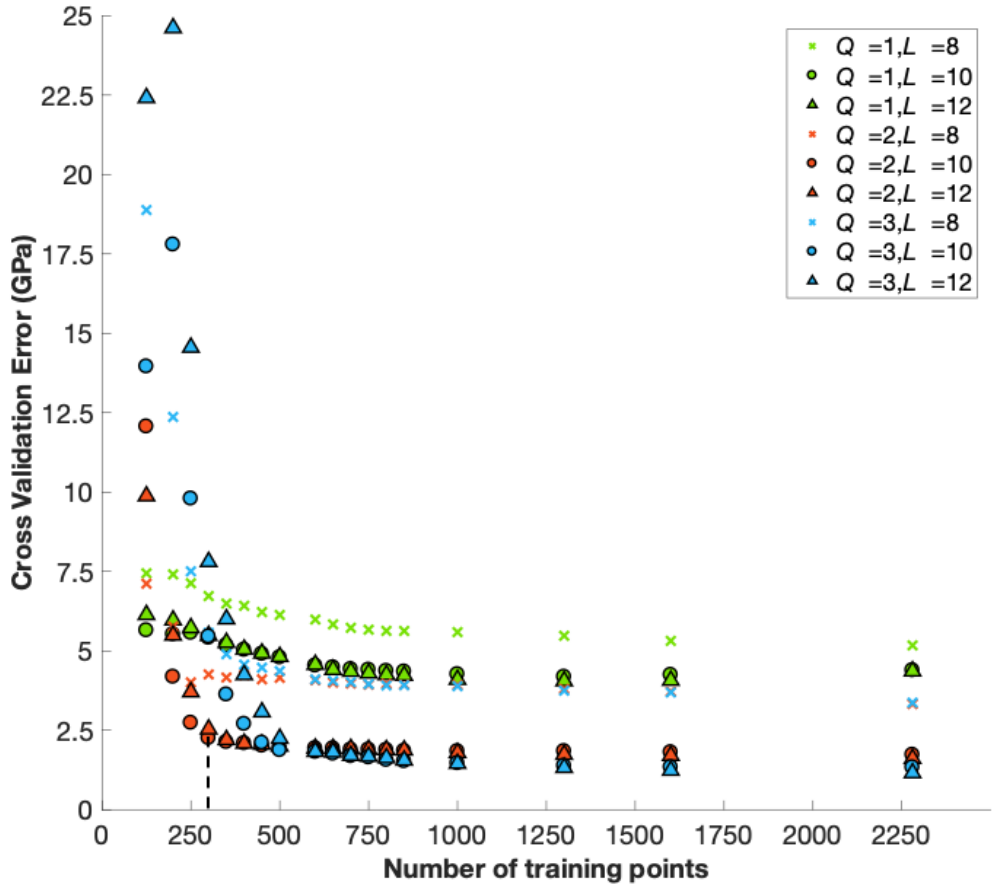

*Figure 1. Cross validation error of the reduced-order model built in the first step of the two-step protocol for different truncation levels in Eq. (1). The dashed line indicates 300 training points.*

For the present study, a set of 123 FE simulations were selected from the previously performed 2286 simulations as this initial set. This initial set was selected to correspond to the boundaries of the intrinsic material parameter space. Following initialization, the reduced-order model in Eq. (1) was considered with different truncation levels of $L$ = 8, 10, 12 for the symmetrized Surface spherical harmonics (differently shaped symbols in Figure 1) and $Q$ = 1, 2, 3 for the maximal degree of the respective Legendre Polynomials [16] (different colors of symbols in Figure 1). The truncation levels of the reduced-order model can be treated as hyperparameters, and must be selected so that we produce the most robust and accurate reduced-order models. Leave-one-out-cross-validation (LOOCV) was performed at various times during the update process and plotted in Figure 1 for the different truncation levels considered.

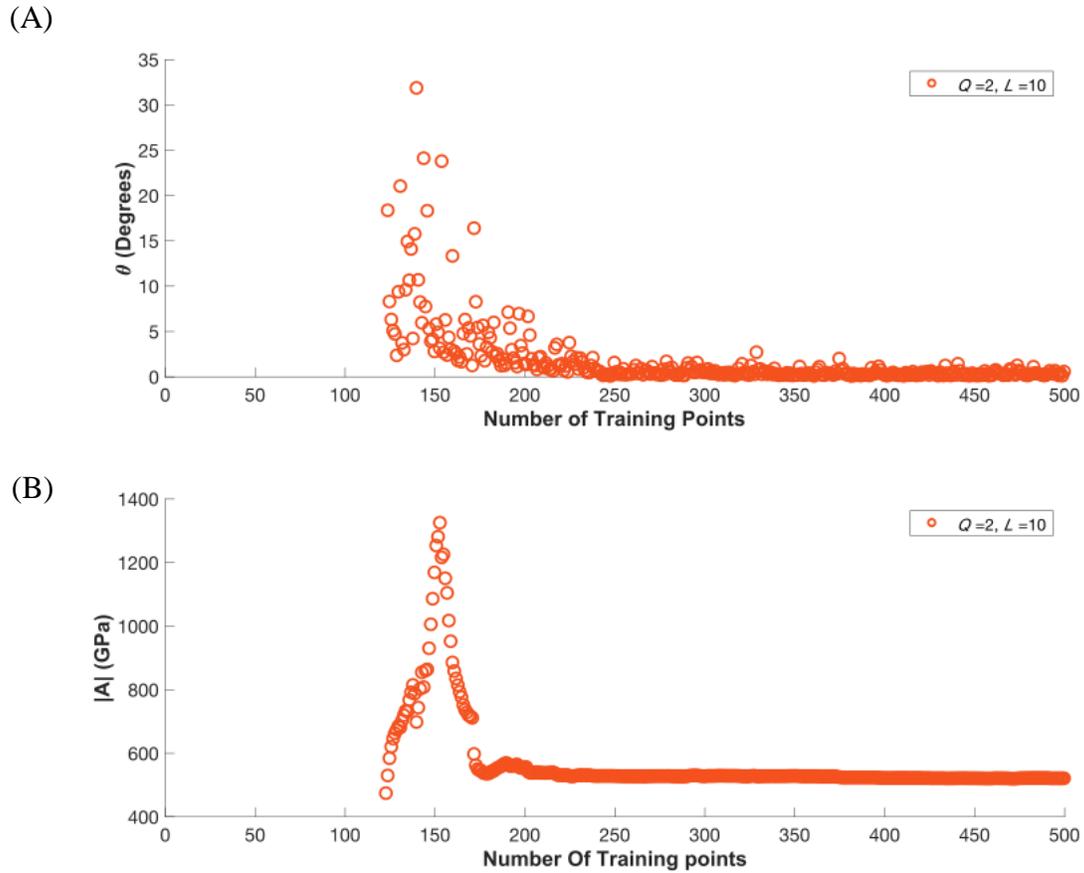

*Figure 2. (A) Variation in the angular change in the vector of model coefficients between model update steps. (B) Variation in the magnitude of the vector of model coefficients during the model building process.*

There is clear improvement in cross validation error up to truncations levels Q=2, L=10, with little improvement for higher truncation levels. The plots in this figure also provide guidance on where to stop the model building effort (i.e., when there is no appreciable improvement in the accuracy of the reduced-order model being built). In addition to the LOOCV, the norm of the vector of model coefficients at each update step (see Figure 2B) and the angular difference of the vector of model coefficients from the previous update step (see Figure 2A) were taken into consideration in determining when to stop the model building effort. Based on these considerations (see Figures 1, 2A, and 2B) it was decided to stop the model building effort after using 300 training points (this includes the set of 123 training point used for initialization). The predictive accuracy of the reduced-order model for the remaining FE simulations (i.e., 2286-300 = 1986) is presented in Figure 3 as a parity plot. The resulting mean absolute prediction error of the reduced-order model was found to be 2.16 GPa (see Figure 3) while the LOOCV error was found to be 2.22 GPa (see

Figure 1). This is comparable to previous efforts based on standard regression techniques and utilizing the full database of 2286 FE simulations, where the LOOCV error was reported to be in the range of 2-2.5 GPa [9].

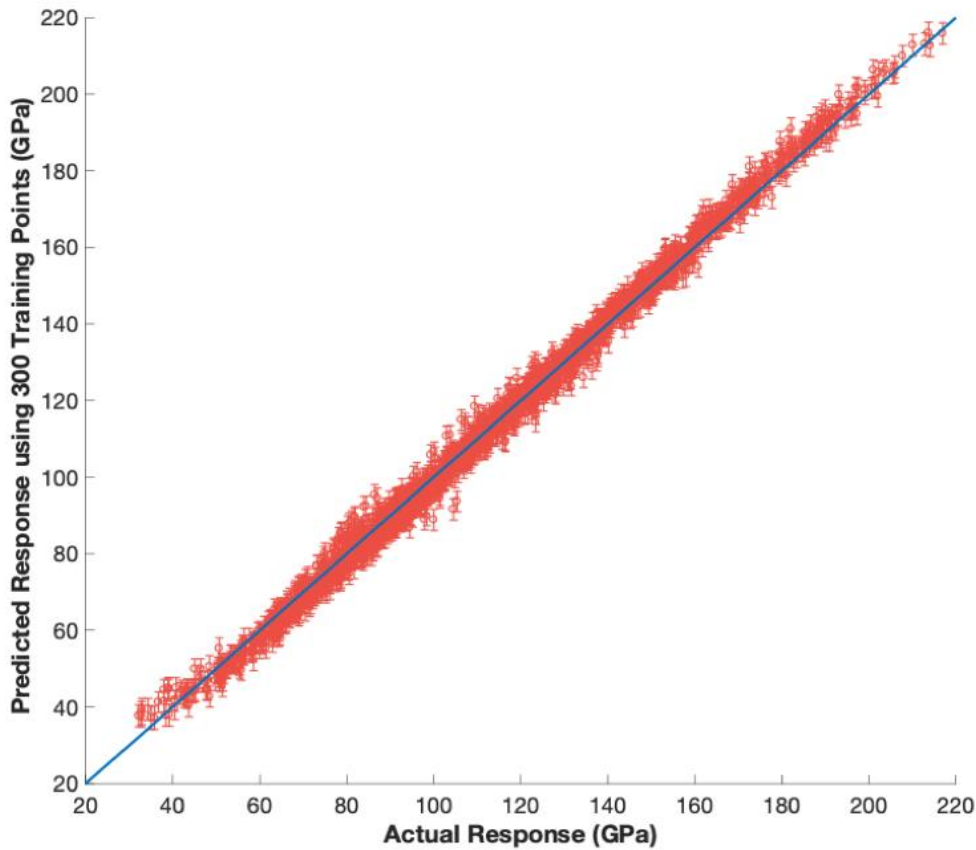

*Figure 3. Predictions for the 1986 FE simulated indentation moduli not used in the training of the reduced-order model. A single standard deviation from the predicted mean is also shown in the plot for each prediction.*

**3.3 Extracting Intrinsic Material Parameters**

At any point during the model building process, the Bayesian framework presented in Section 2.1 can be used to sample the posterior distribution on the material parameters via the MCMC approach. In order to accomplish this second step of the proposed framework, one needs to evaluate the likelihood function (see Eq. (13)); this requires the use of the reduced-order model obtained in step (1) as well as the relevant experimental indentation data. The reduced-order model with truncations $Q=2$, $L=10$ obtained after using 300 training points (described in Section 3.2) was selected for this example case study. Experimental data, including the mean and associated variance of measured indentation moduli, were previously reported for 11 different grains in a polycrystalline sample of Fe 3%-Si [37]. Using the MCMC procedure described in Section 2.2, 50000 samples were drawn. The resulting multivariate distribution is shown in Figure 4 as three univariate distributions.

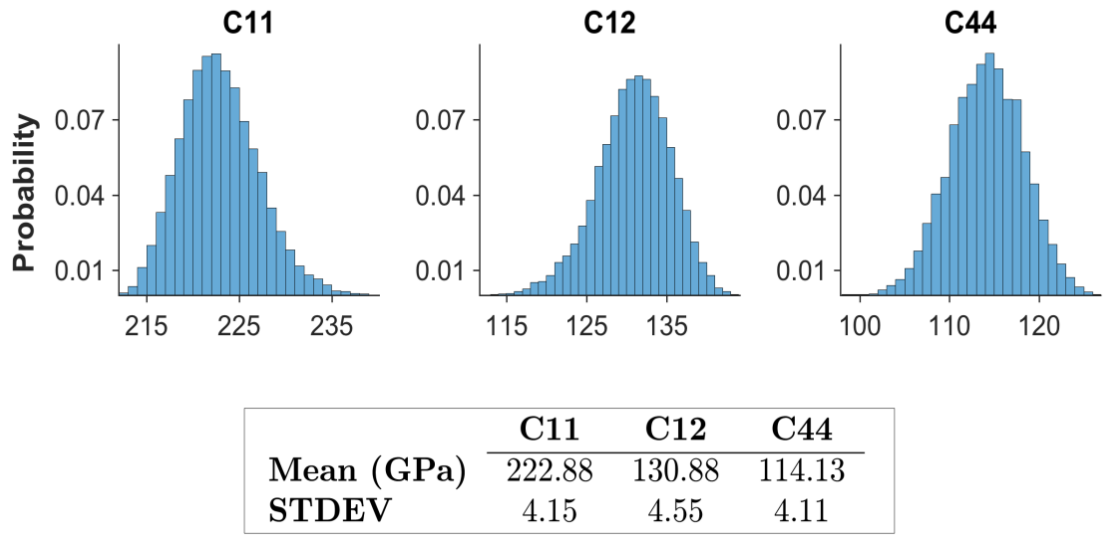

|  | C11 | C12 | C44 |
|---|---|---|---|
| Mean (GPa) | 222.88 | 130.88 | 114.13 |
| STDEV | 4.15 | 4.55 | 4.11 |

*Figure 4. MCMC sampling of the multi-variate posterior distribution of the three intrinsic elastic constants for Fe 3% Si.*

To recap, in Step (1) of the protocol used a minimal number of finite element simulations to establish a high fidelity reduced-order model. Using experimental data previously reported, [37] the established reduced-order model was used to sample the distribution of elastic constants in Step (2) of the protocol. The distributions for the parameters extracted here are in very good agreement with the literature values. Estimates of the elastic constants from the current study, typical values reported from literature [40], and estimates reported from the previous study based on ordinary regression (on the full set of 2286 FE simulations) [9] are shown in Table 1.

|  | $C_{11}$ | $C_{12}$ | $C_{44}$ |
|---|---|---|---|
| Literature[a] | 225 | 135 | 124 |
| Previous Study[b] | 216 | 132 | 122 |
| Current Study | 223 | 132 | 114 |

a. Simmons and Wang [40]
b. Patel et al. [9]

*Table 1. Comparison of reported estimates for single crystal elastic constants of the bcc-metal, Fe-3%-Si. All units are in GPa.*

It is emphasized that the previous study did not attempt any form of uncertainty quantification with respect to these estimates. It is important to note that the highest relative uncertainty in the present study was associated with the estimation of $C_{44}$, which deviated the most from the reported literature values. Since the literature values seldom report the associated uncertainty, it is very difficult to identify the source of the small disagreement between the $C_{44}$ values extracted here from the indentation measurements and the literature values obtained using completely different techniques. This small difference could be attributed to the experimental measurement errors (in both the indentation protocols employed here as well as the more conventional measurement protocols employed in literature). We further note that it should be possible to further refine the methodology presented here (i.e., Step (2) of the protocol) to identify specific additional grain orientations for indentation measurements that might improve specifically the estimates of $C_{44}$ by reducing its variance. Such refinements will be pursued in future work.

## 4. Case Study: Hexagonal Polycrystals

### 4.1 Problem Statement

In order to demonstrate the versatility of the proposed framework, attention is now turned to the extraction of the elastic constants, $\boldsymbol{c} = \{C_{11}, C_{12}, C_{44}, C_{33}, C_{13}\}$, for the hcp metal CP-Ti (commercially pure titanium) [40]. Unlike the previous case study, a database of previously performed FE simulations was not readily available for this case study. Therefore, FE simulations were designed and performed specifically for this study as demanded by the Bayesian inference framework in the Step (1) of the protocol.

The FE model used for this study is the previously validated Finite Element model [9] developed using the commercial software ABAQUS [41]. The sample mesh consisted of 12,610 C3D8 continuum 3-D elements and is shown in Figure 5. The simulated indents were performed using an analytically defined rigid indenter with a tip radius of 16 µm, consistent with the size used in the experiments on single crystal CP alpha-Ti grains reported in literature [1]. The dimensions of the sample mesh were taken as 9.6 µm X 9.6 µm X 4.8 µm. The FE model was validated by comparing simulated indentation moduli to the theoretical values reported by Vlassak and Nix [4] for zinc single crystals $c = \{161.1, 34.2, 38.3, 61.1, 50.3\}\ GPa$ as shown in Figure 5. The comparisons confirm the linear relationship between the indentation load (P) and the elastic indentation depth ($h_e$) raised to a power of 3/2 for hcp single crystals, as predicted by Vlassak and Nix [4] (note that the original Hertz theory [42] is restricted to isotropic materials).

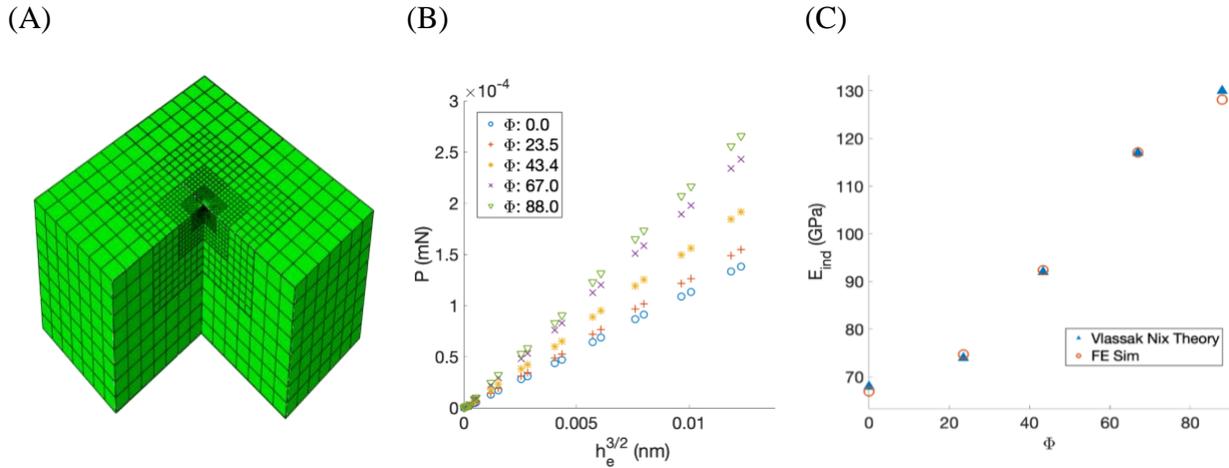

*Figure 5. (A) Mesh used for Finite Element simulations. (B) Finite element simulated plots of P vs $h_e^{3/2}$ for zinc single crystals. (C) Comparison of Theoretical and FE simulated indentation moduli values reported for zinc single crystals.*

For building the reduced-order model (Step 1 of the protocol), we need to identify the specific ranges of the intrinsic material properties of interest. For this study, the bounds of the ranges for the single crystal elastic constants were taken as $80\ \text{GPa} \leq C_{11} \leq 240\ \text{GPa}$, $40\ \text{GPa} \leq C_{12} \leq 120\ \text{GPa}$, $30\ \text{GPa} \leq C_{44} \leq 90\ \text{GPa}$, $70\ \text{GPa} \leq C_{33} \leq 210\ \text{GPa}$, and $40\ \text{GPa} \leq C_{13} \leq 90\ \text{GPa}$; these were chosen to encompass a large number of hcp metals of future interest to our research [6]. The transverse elastic isotropy of the hcp symmetry implies that the elastic indentation response is dependent solely on the declination angle (Φ) between the indenter axis and c-axis of the hcp crystal. Therefore, one only needs to explore the orientation space defined by $0 \leq \Phi \leq$ $\frac{\pi}{2}$ radians. Our goal will be to employ the sequential design strategy once again to efficiently explore the multi-dimensional parameter space identified above in establishing a reliable and robust reduced-order model for the FE indentation simulations over the entire parameter space of interest.

## 4.2 Model Building Process

As with the previous case study, the truncation parameters (Q, L) are important hyper-parameters in the model building process. Since, these are not known a priori, we need to build reduced-order models with different values of these hyper-parameters and make suitable selections. The basic strategy employed here as follows: (1) Reduced-order models with lower truncation levels are initially established, (2) the truncation level is increased systematically if the performance of the established reduced-order model is deemed inadequate, and (3) the model building process is stopped when either the accuracy of the reduced-order model is deemed adequate or when the improvements in the accuracy were deemed insignificant. The LOOCV errors obtained from this process for the different truncations levels are depicted in Figure 6. A set of 760 simulations were used as the initial set for all of these model-building exercises. This number was chosen to be slighter larger than the number of terms in the expansion of Eq. (1) for the case (Q=2, L=4), which results in a total of 729 terms in the expansion. This initial set was identified using a Latin hypercube design (LHD) [43] across the 6 dimensional parameter space $\{C_{11}, C_{12}, C_{44}, C_{33}, C_{13}, \Phi\}$.

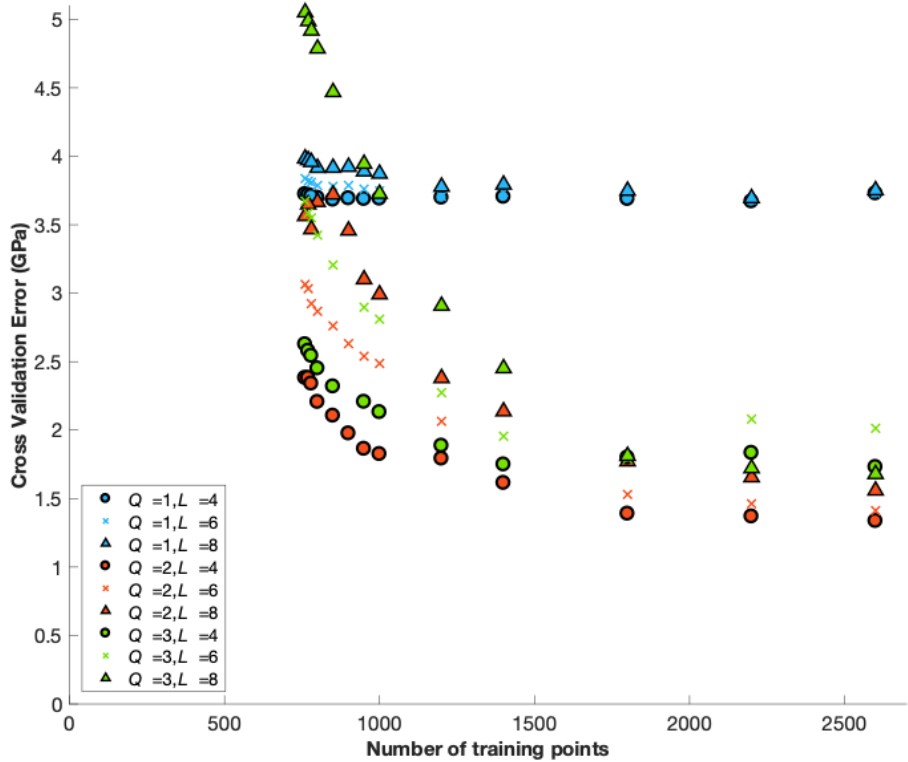

*Figure 6. Cross validation error of the reduced-order models built in the first step of the two-step protocol for different truncation levels of Eq. (1) for hcp crystals.*

Following the initialization, additional simulations were chosen based on a screening of the highest uncertainty across a denser LHD of 2440 sets of inputs (total of 3200 design points including the initialization set). The LOOCV error for the various truncation levels appears to decrease for all cases as data is added with slight increase for the truncations (Q=3, L=4, 6) after 2200 data points, which given the small changes ( less than 0.3 GPa ) is attributed to noise. It is apparent from Figure 6 that the truncation level combination (Q=2, L=4) outperforms others throughout the model building process. The good accuracy of the reduced-order model built for this case study becomes apparent after about 2200 FE simulations, exhibiting a LOOCV error of 1.3 GPa as seen from Figure 6 and Figure 7.

In order to generate a validation set, the selection process was continued to generate another set of 600 FE simulations. We argue that this approach is likely one of the best strategies for building validation sets, as the elements of the validation set are selected based on the highest values of the prediction uncertainty. The prediction errors for the validation set of 600 FE simulations using the reduced-order model built with the training set of 2200 FE simulations are shown in Figure 8. This comparison yields a mean absolute error of 1.2 GPa.

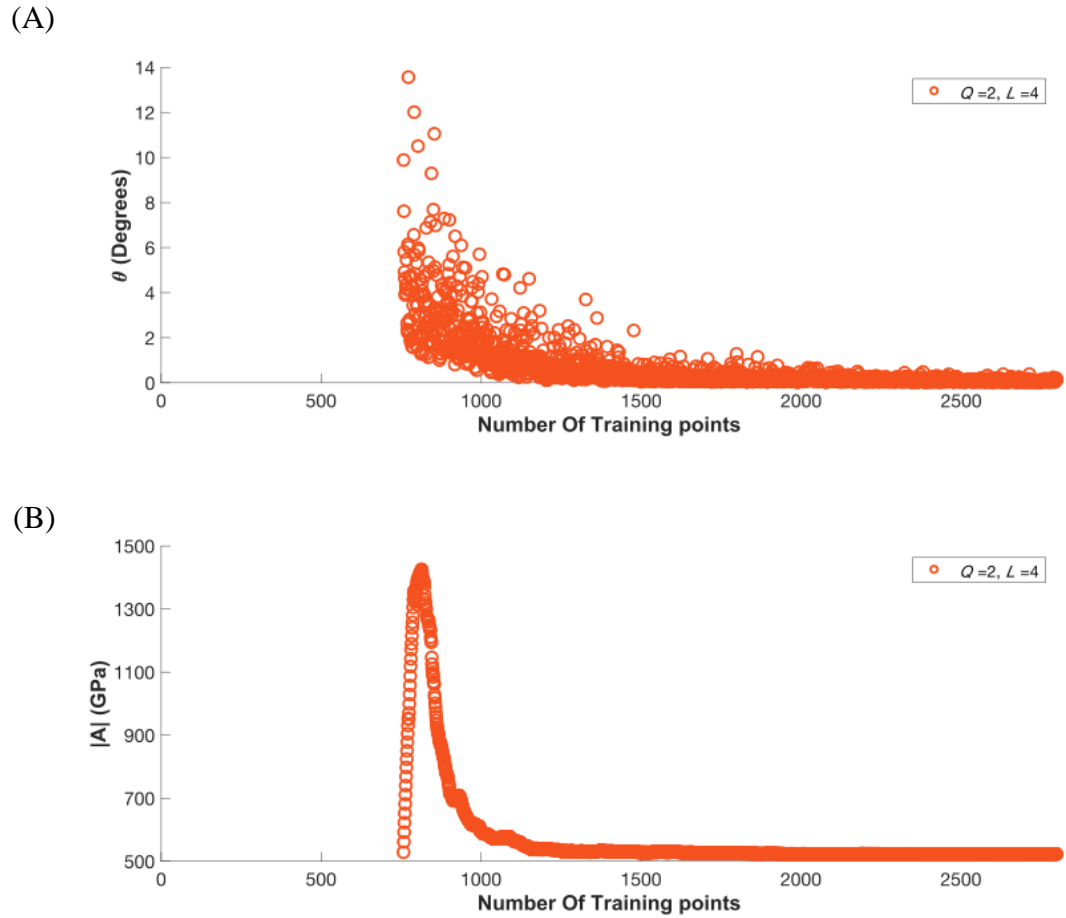

*Figure 7. (A) Variation in the angular change in the vector of model coefficients between model update steps. (B) Variation in the magnitude of the vector of model coefficients during the model building process.*

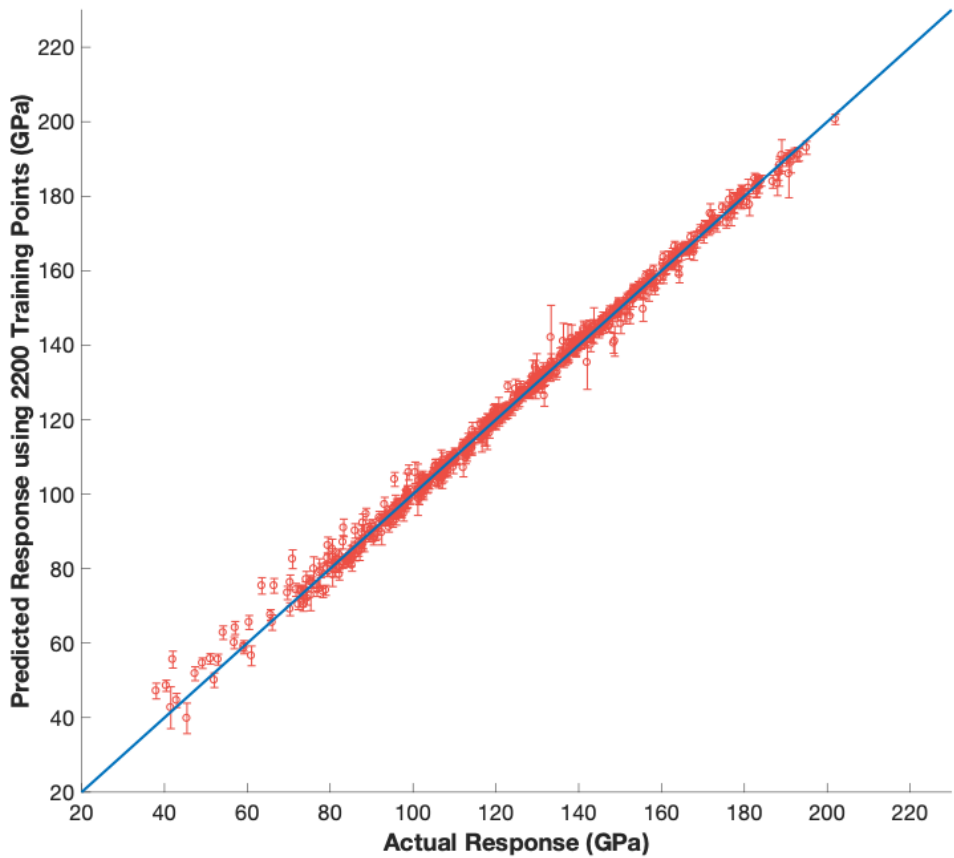

*Figure 8. Predictions for the validation set of 600 FE simulations generated from sequential design process using reduced order model with truncations Q=2,L=4*

It is important to recognize that the parameter space was purposefully chosen to be applicable to many hcp metals of future interest to our research [6]. Predictions are very good over the chosen parameter space as shown in Figure 8. Therefore, within the defined parameter space, future extraction efforts would no longer necessitate the generation of a new model. Furthermore, there is little value in performing additional simulations within the defined parameter space to attempt to significantly improve the reduced-order model. The convergence of the associated model parameters in Figure 7 provides evidence that the reduced-order model is unlikely change drastically with the introduction of new simulations.

It should be noted that the significantly larger training set needed for this case study compared to the previous case study can be attributed to the following reasons: (i) the present case study involved a six-dimensional input space whereas the previous one involved a five-dimensional input space, (ii) the range of values for each input in this case study were selected to be significantly larger than the previous one, and (iii) the degree of elastic anisotropy and contrast captured in this case study is significantly larger compared to the previous case study. The degree of single crystal elastic anisotropy, can be quantified by the universal elastic anisotropic index, $\mathbb{A}$, [44, 45] defined as

$$\mathbb{A} = 5\frac{G_v}{G_r} + \frac{K_v}{K_r} - 6 \tag{14}$$

where $K$ and $G$ are the bulk and shear moduli provided by Voigt and Reuss estimates (indicated by subscript $v$ and $r$ respectively) of a macroscopically homogenous polycrystalline material with uniform texture [46]. A maximum universal elastic anisotropic index of 7.2 was noted for the earlier cubic case study discussed in this paper, compared to 66.2 encountered in the current hcp case study. It is therefore quite reasonable that the number of training data points needed is significantly higher.

### 4.3 Extracting Intrinsic Material Parameters

The focus is now turned to the sampling of the posterior distribution of the elastic constants, $\boldsymbol{c} = \{C_{11}, C_{12}, C_{44}, C_{33}, C_{13}\}$, via MCMC. Similar to the previous case study, in order to sample from the posterior distribution of the intrinsic material parameters, the likelihood function in Eq. (13) must be computed using the available experimental data and the reduced-order model established in Step (1) (corresponding to truncation levels Q=2, L=4 using a training set of 2200 FE simulation data points). The experimental data for this case study was obtained from a prior openly shared dataset [1]. This data set included indentation moduli for 50 different crystal orientations on a CP-Ti sample. Following the procedure described in Section 2.2, 50000 samples were drawn using the MCMC approach. The resulting posterior distributions are shown in Figure 9 for each of the five intrinsic hcp elastic stiffness parameters. The maximum-a-posteriori (MAP) estimates, the mean values, and the standard deviations of the distribution are reported as a table in the same figure. The reported mean values for elastic constants were found to be $\{155, 89, 49, 174, 55\}$ GPa for $\{C_{11}, C_{12}, C_{44}, C_{33}, C_{13}\}$, respectively. Typical literature values reported are $\{162, 92, 47, 180, 69\}$ GPa [32]. With the exception of $C_{13}$, the extracted intrinsic stiffness parameters show good agreement with values reported in literature (mean values are within 5%). It is also interesting to note that the extracted distribution for $C_{13}$ exhibits the highest relative uncertainty.

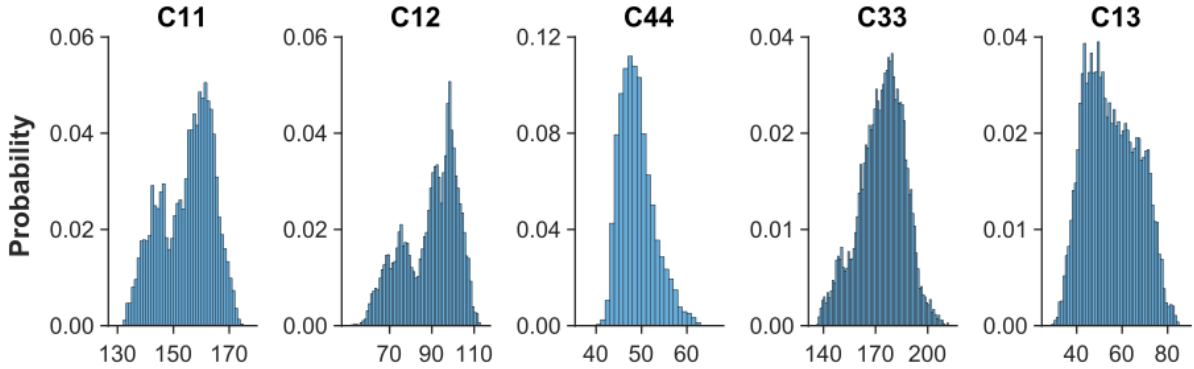

*Figure 9. MCMC sampling of the posterior distribution for the intrinsic single crystal elastic constants of CP-Ti.*

This indicates the relative low sensitivity of the indentation modulus to changes in $C_{13}$, when compared to the other elastic stiffness constants. As noted in the previous case study, it should be possible to extend the framework presented here to focus exclusively on improving the estimation of $C_{13}$ [13]. However, such an effort could only be justified after the uncertainty in the literature reported values is rigorously quantified. The variance in the predictions of the surrogate model at selected orientations is compared with the experimental data in Figure 10.

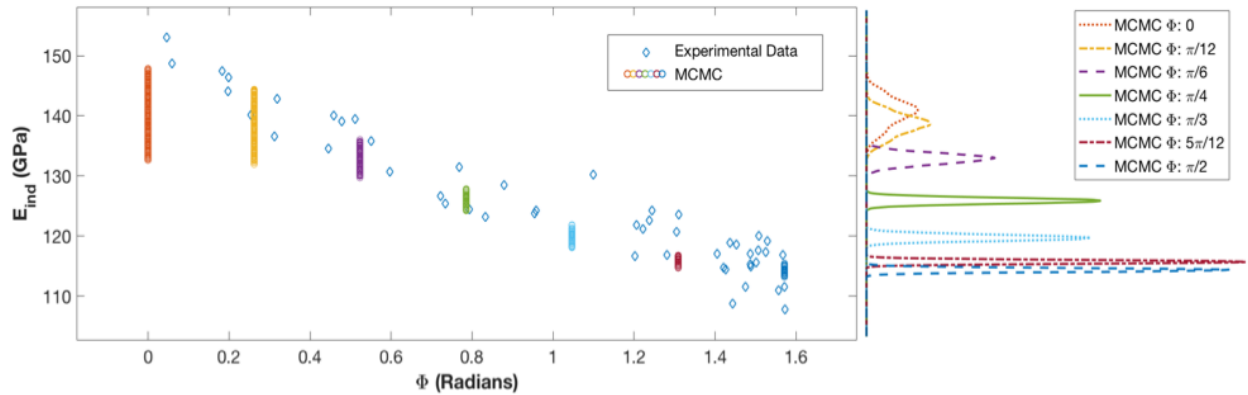

*Figure 10. Left: Reduced-order model evaluations of the Markov Chain (MCMC) at selected points across the orientation space compared to available experimental data. Right: The resulting distributions of the evaluations using the reduced order model.*

Evaluations of the reduced-order model at various orientations using samples from the posterior distributions of the elastic constants provides the possible mean indentation moduli for the observed experimental indentation moduli, as described in Section 2.2. Since the reduced-order model coupled with a sampled set of elastic moduli from the posterior distribution of elastic constants provides the respective mean indentation modulus as a function of orientation, the predictions should be more tightly packed in regions which there are more observations, reflecting a higher certainty of the mean. The prediction uncertainty from MCMC is in fact shown to be highest at low declination angles, while uncertainty is lowest at high declination angles where relatively much more data is available. Furthermore, this observation suggests that there is much more value in conducting additional tests at the lower declination angles, specifically in the range of 0-0.2 radians, compared to conducting them at the higher declination angles. This could be highly valuable input to the experimentalists for their future studies.

## 5. Conclusions

A statistical framework has been presented for the robust extraction of the intrinsic material parameters from available experimental observations from spherical indentation stress-strain protocols. The two-step Bayesian inference framework enables the specification of uncertainty in the measurement data, which is then transferred to the uncertainty in the values of the extracted intrinsic material properties. Most importantly, the new framework presented in this paper demonstrates potential for significantly speeding up the materials characterization effort by focusing on experiments that are likely to deliver the maximum value in establishing the desired properties. This is accomplished by employing a numerical model of the experiment itself (here accomplished using a finite element model). Although the numerical model can be very expensive, it is only needed for a one-time effort is establishing a reduced-order model (Step (1) of the proposed two-step protocol). Once the reduced-order model is established, the calibration of the available experimental data to the theory (Step (2) of the proposed two-step protocol) can be accomplished with minimal computational resources. The versatility and the robustness of the proposed new framework is demonstrated with two case studies: (i) extraction of three elastic constants for Fe-3%-Si, and (ii) extraction of the five elastic constants for CP-Ti. In both case studies, the ranges of intrinsic material parameters considered covers a significant number of polycrystalline hcp and cubic metals. This makes both models highly applicable to new case studies within the material classes. For material classes outside of the classes explored here, the main challenge is indeed Step (1) of the protocol, requiring the establishment of a high fidelity reduced-order model from suitable FE simulations, while Step (2) remains the same. In the event the extracted parameters in Step (2) fall outside of the extents of the databases used to construct the reduced-order model, additional simulations considering the new bounds would become necessary. Finally the use of a Bayesian framework opens new avenues for the development of autonomous (fully guided by the computer) scientific explorations. It is anticipated that the framework is extensible to a large number of other applications in multiscale materials modeling (e.g., extraction of the values of slip resistances from indentation measurements, extraction of the values of parameters in phase-field models based on available microstructure datasets).

**Correspondence:** Surya R. Kalidindi, surya.kalidindi@me.gatech.edu

**Conflicts of Interest**

The authors declare that the research was conducted in the absence of any commercial or financial relationships that could be construed as a potential conflict of interest.

**Funding**

The authors acknowledge funding from AFOSR award FA9550-18-1-0330 (Program Manager: J. Tiley). AC acknowledges funding from the National Science Foundation Grant 1258425.

main challenge is indeed Step (1) of the protocol, requiring the establishment of a high fidelity reduced-order model from suitable FE simulations, while Step (2) remains the same. In the event the extracted parameters in Step (2) fall outside of the extents of the databases used to construct the reduced-order model, additional simulations considering the new bounds would become necessary. Finally the use of a Bayesian framework opens new avenues for the development of autonomous (fully guided by the computer) scientific explorations. It is anticipated that the framework is extensible to a large number of other applications in multiscale materials modeling (e.g., extraction of the values of slip resistances from indentation measurements, extraction of the values of parameters in phase-field models based on available microstructure datasets).

**Correspondence:** Surya R. Kalidindi, surya.kalidindi@me.gatech.edu

**Conflicts of Interest**

The authors declare that the research was conducted in the absence of any commercial or financial relationships that could be construed as a potential conflict of interest.

**Funding**

The authors acknowledge funding from AFOSR award FA9550-18-1-0330 (Program Manager: J. Tiley). AC acknowledges funding from the National Science Foundation Grant 1258425.

6. Priddy, M.W., *Exploration of forward and inverse protocols for property optimization of Ti-6Al-4V*, D.L. McDowell, et al., Editors. 2016, Georgia Institute of Technology.
7. Patel, D. and S. Kalidindi, *Estimating the slip resistance from spherical nanoindentation and orientation measurements in polycrystalline samples of cubic metals.* International Journal of Plasticity, 2017. **92**: p. 19.
8. Sánchez-Martín, R., et al., *Measuring the critical resolved shear stresses in Mg alloys by instrumented nanoindentation.* Acta Materialia, 2014. **71**: p. 283-292.
9. Patel, D.K., H.F. Al-Harbi, and S.R. Kalidindi, *Extracting single-crystal elastic constants from polycrystalline samples using spherical nanoindentation and orientation measurements.* Acta Materialia, 2014. **79**: p. 108-116.
10. Zambaldi, C., et al., *Orientation informed nanoindentation of α-titanium: Indentation pileup in hexagonal metals deforming by prismatic slip.* Journal of Materials Research, 2012. **27**(1): p. 356-367.
11. Bhattacharya, A.K. and W.D. Nix, *Finite element simulation of indentation experiments.* International Journal of Solids and Structures, 1988. **24**(9): p. 881-891.
12. Gelman, A., et al., *Bayesian Data Analysis, Third Edition (Chapman & {Hall/CRC} Texts in Statistical Science)*. 2014: Chapman and Hall/CRC.
13. Huan, X. and Y.M. Marzouk, *Simulation-based optimal Bayesian experimental design for nonlinear systems.* Journal of Computational Physics, 2013. **232**(1): p. 288-317.
14. MacKay, D.J.C., *Introduction to Gaussian process.* Neural Networks and Machine Learning, 1998.
15. Rasmussen, C.E., *Evaluation of Gaussian processes and other methods for non-linear regression*. University of Toronto.
16. Bunge, H.-J., *Texture analysis in materials science. Mathematical Methods*. 1993, Göttingen: Cuvillier Verlag.
17. Adams, B.L., S.I. Wright, and K. Kunze, *Orientation imaging: the emergence of a new microscopy.* Metallurgical Transactions A (Physical Metallurgy and Materials Science), 1993. **24A**(4): p. 819-31.
18. Adams, B.L., S.R. Kalidindi, and D.T. Fullwood, *Microstructure sensitive design for performance optimization*. 2012, Elsevier Science: Oxford.
19. Proust, G. and S.R. Kalidindi, *Procedures for construction of anisotropic elastic-plastic property closures for face-centered cubic polycrystals using first-order bounding relations.* Journal of the Mechanics and Physics of Solids, 2006. **54**(8): p. 1744-1762.
20. Knezevic, M., S.R. Kalidindi, and R.K. Mishra, *Delineation of first-order closures for plastic properties requiring explicit consideration of strain hardening and crystallographic texture evolution.* International Journal of Plasticity, 2008. **24**(2): p. 327-342.
21. Yabansu, Y.C. and S.R. Kalidindi, *Representation and calibration of elastic localization kernels for a broad class of cubic polycrystals.* Acta Materialia, 2015. **94**: p. 26-35.
22. Yabansu, Y.C., D.K. Patel, and S.R. Kalidindi, *Calibrated localization relationships for elastic response of polycrystalline aggregates.* Acta Materialia, 2014. **81**: p. 151-160.
23. Box, G.E.P., *Bayesian inference in statistical analysis*, ed. G.C. Tiao. 1973, Reading, Mass.: Reading, Mass., Addison-Wesley Pub. Co.
24. Gelman, A., *Bayesian data analysis*. 2nd edition.. ed. 2004, Boca Raton, Fla.: Boca Raton, Fla. : Chapman & Hall/CRC.